\documentclass[twocolumn]{aa}
\usepackage{natbib}
\usepackage{color}
\usepackage{ragged2e}
\usepackage[pdfpagelabels=false]{hyperref}	
\hypersetup{colorlinks=true,linkcolor=blue,citecolor=blue,filecolor=blue,urlcolor=blue,}
\usepackage[varg]{txfonts}
\usepackage{graphicx,rotating}
\usepackage[normalem]{ulem}
\usepackage{colortbl}
\usepackage{xcolor}
\usepackage{bbold}
\usepackage{upgreek}

\bibpunct{(}{)}{;}{a}{}{,} 

\definecolor{darkolivegreen}{rgb}{0.33, 0.42, 0.18}

\definecolor{asparagus}{rgb}{0.53, 0.66, 0.42}

\definecolor{salmon}{rgb}{0.95,0.5,0.25}

\newcommand{\lcdm}{$\Lambda$CDM }
\newcommand{\Msol}{\mathrm{\, M}_\odot}
\newcommand{\kpc}{\mathrm{\, kpc}}

\def\e{\rm{e}}
\def\RG{R_{\rm G}}
\def\Mpch{\,\mathrm{Mpc}/h}
\def\Mpc{\,\mathrm{Mpc}}
\def\nuf{\nu_{\rm f}}
\def\Msolh{\,\mathrm{M_\odot}/h}

\begin{document}

\title{Distinguishing cold and self-interacting dark matter through topological analysis}

\titlerunning{Distinguishing cold and self-interacting dark matter}
\author{Adrian Szpilfidel \inst{1,2}, Clotilde Laigle \inst{1}, Pierre Boldrini \inst{3}, Moritz S.\ Fischer \inst{4,5} and Dmitri Pogosyan \inst{6}}
\offprints{Adrian Szpilfidel, \email{adrian.szpilfidel@obspm.fr }}
\institute{$^{1}$ LUX, Observatoire de Paris, Université PSL, Sorbonne Université, CNRS, 92190 Meudon, France\\$^{2}$ Institut d’Astrophysique de Paris, UMR 7095, CNRS, Sorbonne
Université, 98 bis boulevard Arago, F-75014 Paris, France \\ $^{3}$ LIRA, Observatoire de Paris, Université PSL, Sorbonne Université, Université Paris Cité, CY Cergy Paris Université, CNRS, 92190 Meudon, France \\ 
$^{4}$ Donostia International Physics Center (DIPC), Paseo Manuel de Lardizabal 4, 20018 Donostia-San Sebastián, Spain \\
$^{5}$ Universitäts-Sternwarte, Fakultät für Physik, Ludwig-Maximilians-Universität München, Scheinerstr. 1, 81679 München,
Germany \\ $^{6}$ Physics Department, University of Alberta, 4-181 CCIS, Edmonton AB T6G 2E1,  Canada}
\authorrunning{Szpilfidel et al.}
\date{submitted to A$\&$A}

\abstract{Alternative dark matter (DM) models have emerged to solve the challenges faced by the predictions of collisionless cold dark matter (CDM) on galactic scales ($\lesssim 1\Mpc$). However, disentangling alternative models from CDM is difficult on such small scales because of the degeneracy with baryonic physics. It is therefore necessary to use DM probes that are not affected by baryons, e.g.\ that stand on intermediate scales, larger than galactic while remaining smaller than the scale at which the models converge to CDM.
For the first time, we distinguish self-interacting DM (SIDM) from CDM using the genus statistic, a metric that characterises the topology of the density field. 
We carried out the analysis on the Darkium DM-only cosmological simulations, using one CDM model and four SIDM models with cross-sections of various amplitudes and velocity dependencies. We computed the genus on selected 3-virial radius wide regions centred around halos, for few hundred halos with masses ranging from $10^{12}$ to $10^{14}$ M$_\odot/h$ over redshifts $z=0$ to $z=2$. We also explored a more observation-like configuration, where the DM density field is traced only from the halo distribution in thick 2D projection since in principle redshift errors hinder a 3D reconstruction of the density field. 
We find that the density field is systematically clumpier in CDM than in SIDM models up to $0.05~\mathrm{Mpc}/h$, for halos of masses larger than $10^{12}\Msol/h$ at $z=0$.
These predictions show that the genus of the density field is sensitive to DM self-interactions, suggesting that topological analysis could provide a valuable probe for distinguishing SIDM from CDM in observed halo distributions.
}

\maketitle



\section{Introduction}

\lcdm has become the standard cosmological model describing the evolution of the mass and energy content of the Universe, following the discovery of the accelerated cosmic expansion by \cite{riess_observational_1998}. Two major ingredients of the \lcdm model are the uniform dark energy and the matter that is able to undergo gravitational collapse. Current measurements indicate that 82\% of the matter component of the Universe is made of an invisible species that interacts only through gravity \citep{planck_collaboration_planck_2020}. This dark matter (DM) component is usually described as a collisionless fluid, with low velocity dispersion, known as Cold Dark Matter (CDM). The remaining 18\% includes ordinary baryonic matter. Within the \lcdm framework, CDM has not been associated with any known elementary particle, and its nature remains an open question.  \lcdm has been remarkably successful in reproducing observations at cosmological scales ($\gtrsim10\Mpc$), where simulations and large-scale surveys show excellent agreement with predictions \citep{eisenstein_detection_2005, beutler_clustering_2017, planck_collaboration_planck_2020, alam_completed_2021}. However, on smaller, galactic scales ($1$--$100~\kpc$), tensions remain between theoretical predictions and observations. Dwarf galaxies often exhibit central cores rather than the steep cusps predicted by CDM: the cusp–core problem \citep[e.g.][]{blok_core-cusp_2010,boldrini_cusp-core_2021}. The CDM cusps were found in cosmological simulations predicting a density profile that diverges as $\rho \propto r^{-1}$ towards the centre of DM halos \citep{navarro_universal_1997}. In contrast, observations of dwarf galaxies suggest a flat density profile in their central regions \citep{moore_evidence_1994,de_blok_high-resolution_2001}. The number of galaxy satellites is of great interest in comparing simulations to observed satellites \citep[the missing‑satellites problem,][]{klypin_where_1999, Kim_2018, Kanehisa_2024}, and Milky Way–like halos appear to host subhalos that are too dense compared to data \citep[the too‑big‑to‑fail problem,][]{boylan-kolchin_too_2011}, and the debate around the diversity of rotation curves \citep{2015MNRAS.452.3650O, raman2026semianalyticinferencesatellitedensities, cruz2025dwarfdiversitylambdacdmbaryons}. For a comprehensive review of small‑scale challenges to CDM, see \cite{bullock_small-scale_2017}.

These discrepancies have motivated the exploration of alternative models. A promising scenario for solving these problems introduces DM with self-interactions, known as Self-Interacting Dark Matter (SIDM), first developed by \cite{carlson_self-interacting_1992} and \cite{spergel_observational_2000}. SIDM has later been investigated through cosmological simulations \citep[e.g.][]{rocha_cosmological_2013, Robertson_2019, correa_tangosidm_2022, despali_introducing_2025}. This model introduces a new free parameter for DM self-interactions, the cross-section $\sigma_0$. SIDM behaves like CDM in the limit of null cross-section. However, assuming a non-zero cross-section leads to a different DM distribution on galactic scales. \cite{dave_halo_2001, rocha_cosmological_2013,elbert_core_2015} showed that SIDM can produce cored density profiles in dwarf galaxies through effective heat conduction, alleviating the cusp--core problem: the SIDM particles in the centre of the halo gain kinetic energy via self-interactions, leading to more extended orbits and, therefore, a reduction in central density. This process redistributes energy, preventing the formation of gravitationally bound DM clumps, resulting in more diffuse matter distribution. \cite{zavala_constraining_2013,kaplinghat_too_2019} demonstrated that the too-big-to-fail' problem might be solved by SIDM. Self-interactions result in a reduced number of satellites, in particular with small-angle scattering \citep{fischer_cosmological_2022, 2026arXiv260319362K}. For a review of SIDM models and their implications on small-scale structures, see \cite{tulin_dark_2018, adhikari_2025}. SIDM can provide a diversity of halos \citep{Ren_2019, nadler_symphony_2023} that is in better agreement with observations on galactic scales while remaining consistent with large-scale structure formation.

In practice, obtaining robust constraints on DM self-interactions from galaxy cores is extremely challenging: complex baryonic physics in these regions, such as bars, gas inflows and outflows, and bursty stellar feedback, redistribute mass and energy and can mimic or erase SIDM signatures. A more effective strategy is to probe larger scales, which mitigate baryonic degeneracies while still lying in a regime where SIDM gravitational heating yields smoother and more diffuse structures. This is precisely where topological diagnostics can help distinguish SIDM from CDM. On scales larger than few$\Mpc$s, matter is distributed along a cosmic web \citep[][]{bond_how_1996, pogosyan_cosmic_1998}. This complex network consists of filaments and walls, intersecting at nodes, surrounding low-density regions (voids). The DM distribution, being highly anisotropic at these scales, induces gravitational potential wells that drive the infall of baryonic matter along filaments towards their intersections (nodes), where halos grow, shaping the large-scale structure of the Universe. As gas accretes into a halo, it radiatively cools, loses pressure support, and fragments \citep{rees_cooling_1977,white_galaxy_1991, white_core_1978, mo_galaxy_2010}. This triggers star formation \citep{barkana_beginning_2001}, building the first galaxies in the densest knots of the cosmic web. Therefore, the observed galaxy distribution is a powerful tracer for probing the underlying DM distribution \citep{davis_survey_1982, de_lapparent_slice_1986, geller_mapping_1989}. Measuring the geometry of the cosmic web provides insight into its growth and content. Among the various existing methods for studying the topology of the cosmic web, one approach is to use the genus of the density contours \citep{gott_quantitative_1987, park_topology_2005}, which quantifies the connectivity of the density field. The gravitational heating induced by self-collisions in SIDM, limiting structure formation, might lead to different topology than CDM. Thus, the genus is expected to be sensitive to these topological differences. This method has the advantage of being non-parametric, thereby limiting biases in the analysis. This statistical tool has been used widely in cosmology, for instance to constrain cosmological parameters \citep{appleby_cosmological_2018} or on alternative DM models \citep{watts_large-scale_2017}.

In this work, we compare the topology of the cosmic web formed in CDM and SIDM cosmologies using DM-only cosmological simulations. We investigate the topological properties of DM density distribution in both CDM and SIDM frameworks comparing the results on different parameter sets for SIDM. We also investigate how the differences between the two models evolve with redshift. This comparison aims to assess how DM self-interactions modify the connectivity of large-scale structures and how such effects could be tested with upcoming data from large-scale surveys such as the Euclid mission. The Euclid survey aims to probe the nature of dark energy and DM, it provides high-resolution optical imaging, as well as near-infrared imaging and spectroscopy, over about 14,000 deg$^2$ of extragalactic sky \citep{collaboration_euclid_2025}. With its unprecedented volume and depth, Euclid will map the large-scale structure of the Universe with high precision, making it an ideal dataset for testing cosmological models, including SIDM.

This paper is organised as follows. In \autoref{Method}, we present the genus statistic and how it is applied to the analysis of cosmological simulations. In \autoref{Results}, we report the results of the comparison of the cosmic web topology in CDM and SIDM frameworks. Finally, we summarise and discuss possible perspectives on this work.
Additional information is provided in the appendices.

\section{Methodology}
\label{Method}

This section presents the methodology adopted to compare the DM models. We first describe the cosmological simulations and the DM models, before detailing the analysis based on the genus statistic.

\subsection{Cosmological simulation data}

The analysis presented in this work has been carried out using data 
from cosmological DM-only simulations adopting either CDM or SIDM\footnote{https://www.darkium.org/} \citep{fischer_cosmological_2022, fischer_2024a} paradigms, based on the code \textsc{OpenGadget3} (Dolag et al.\ in prep.)
, which is a successor of \textsc{GADGET-2} \citep{springel_cosmological_2005}.
The domain decomposition and the neighbour search we
used have been described by \cite{ragagnin_2016} and the implementation of SIDM was introduced by \cite{fischer_n_2021, fischer_2022, fischer_cosmological_2022, fischer_2024a, fischer_2026}.

Starting at $z=60$, the simulation ran over a Hubble time ($ 13.8$ Gyr) in a cubic box of comoving size 48 $\Mpch$, with $576^3$ DM particles, with a mass resolution of $4.37\times10^7 \mathrm{M_{\odot}} / h$. The gravitational softening length is set to a comoving value of $4.2 \, \mathrm{ckpc
} / h$ and capped at a maximum physical length of $1.4 \, \mathrm{kpc} / h$.
We employed a cosmological model described by the following parameters: $\Omega_{\mathrm{M}} = 0.272$, $\Omega_\Lambda = 0.728$, $h = 0.704$, $n_\mathrm{s} = 0.963$, and $\sigma_8 = 0.809$ \citep[WMAP7;][]{komatsu_2011}.
The simulations were run with five different DM models, one collisionless and four self-interacting ones. The details can be found in Table~\ref{tab:dm_models}. We note that all simulations started from the same initial conditions.

\begin{table}[]
    \caption{Overview of simulated DM models.}
    \centering
    \begin{tabular}{lcc}
         Name & $\sigma_0$ & $\omega_0$ \\
         & [cm$^2$~g$^{-1}$] & [km~s$^{-1}$] \\ \hline
         CDM & 0.0 & -- \\
         SIDM$\sigma 02\omega\infty$ & 0.2 & $\infty$ \\
         SIDM$\sigma 2\omega\infty$ & 2.0 & $\infty$ \\
         SIDM$\sigma 20\omega 180$ & 20.0 & 180.0 \\
         SIDM$\sigma 200\omega 180$ & 200.0 & 180.0
    \end{tabular}
    \label{tab:dm_models}
    \tablefoot{The first column gives the name of the DM model as used throughout the paper. The second and third columns specify the model parameters $\sigma_0$ and $\omega_0$ of Eq.~\eqref{eq:veldep}. 
    We note that all the SIDM models are in the limit of a very anisotropic cross-section.}
\end{table}

The SIDM models introduce self-interactions between DM particles that alter the matter distribution on small scales. They are typically characterised by a cross-section divided by the physical particle mass.  
To quantify the strength of the self-interactions, we use the viscosity cross-section normalised such that it matches to the total cross-section for isotropic scattering,
\begin{equation} \label{eq:viscosity_cross_section}
\sigma_\mathrm{V} = \frac{3}{2} \int_{-1}^{1} \frac{\mathrm{d} \sigma}{\mathrm{d} \cos \theta_{\mathrm{cms}}}\sin^2\theta_{\mathrm{cms}} \, \mathrm{d} \cos \theta_{\mathrm{cms}} \, .
\end{equation}
Here, $\mathrm{d}\sigma / \mathrm{d}\cos \theta_\mathrm{cms}$ is the differential cross-section, with $\theta_\mathrm{cms}$ being the scattering angle in the centre-of-mass system of the two scattering particles.

In our simulations, we use the so-called frequently self-interacting dark matter (fSIDM) model, which corresponds to highly anisotropic scattering, in other words, small-angle scattering \citep{fischer_n_2021}. It means that the DM particles are slightly deviated, with a small angle from their initial trajectory during an interaction. Consequently, the number of scatter events has to be high to have a significant effect on the DM distribution. This is why this model is called ``frequent''. This SIDM model is motivated by light mediator models that can give rise to highly anisotropic scattering \citep[e.g.][]{kahlhoefer_2014}.

\begin{figure}
    \centering
    \includegraphics[width=1\linewidth]{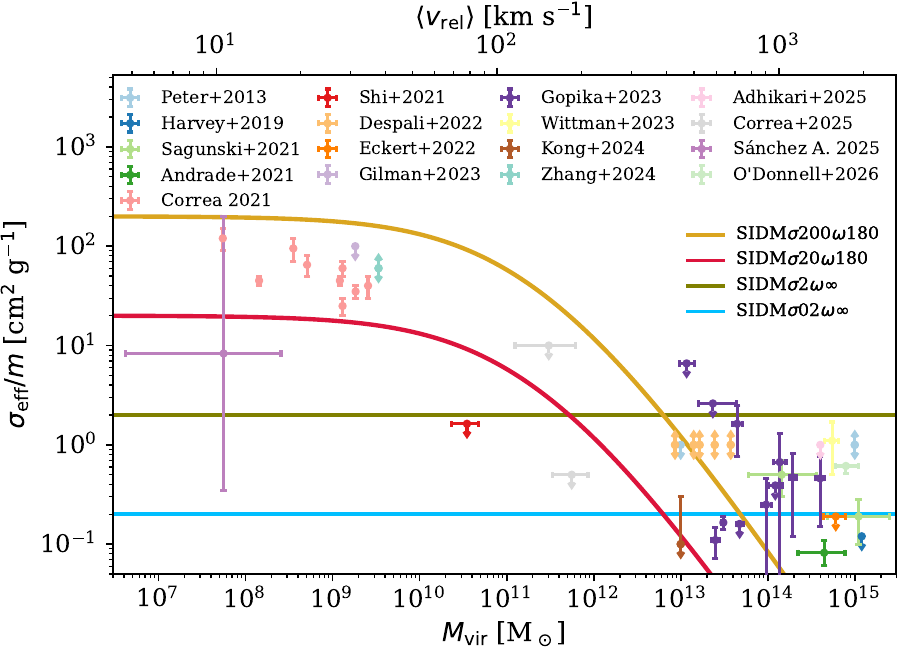}
    \caption{The SIDM models that we study together with several constraints on DM self-interactions. The effective cross-section \citep{yang_2022} for our four models, as given in Table~\ref{tab:dm_models}, is shown as a function of velocity and halo mass \citep[analogously to fig.~G1 by][]{fischer_2024a}. 
    In addition we show several constraints on SIDM derived from a variety of observables \citep{peter_2013, harvey_2019, sagunski_2021, andrade_2021, correa_2021, shi_2021, despali_2022, eckert_2022, gilman_2023, gopika_2023, wittman_2023, kong_2024, zhang_2024, adhikari_2025, correa_2025, sanchez_almeida_2025, odonnell_2026}.}
    \label{fig0}
\end{figure}

Despite the angular dependence, a differential cross-section can also depend on the relative velocity, $v$, of the interacting particles. In our simulations, we use a simple model for the velocity dependence of the cross-section; the viscosity cross-section (Eq.~\eqref{eq:viscosity_cross_section}) per physical particle mass, $m_\chi$, is parametrised as
\begin{equation} \label{eq:veldep}
    \frac{\sigma_\mathrm{V}}{m_\chi} = \sigma_0 \, \left(1+ \left( \frac{v}{\omega_0} \right)^2 \right)^{-2} \,.
\end{equation}
The two parameters of our model are the cross-section normalisation $\sigma_0$ and $\omega_0$ controlling the velocity at which the cross-section transitions from the velocity-independent regime at low velocities to the $\sigma_\mathrm{V} \propto v^{-4}$ regime.
We explore two velocity-independent models and two velocity-dependent models as specified in Table~\ref{tab:dm_models} and illustrated in Fig.~\ref{fig0}.

SIDM particles exchange kinetic energy through collisions. For isolated halos, this can be effectively described by heat conduction \citep[e.g.][]{balberg_2002}. Hence, self-interactions heat the relatively cold cores of DM halos. The speed increase in the central region of the halo causes particles to move to more extended orbits, resulting in a more diffuse DM distribution compared to the collisionless CDM predictions.
The larger cross-sections (SIDM$\sigma$2$\omega\infty$, SIDM$\sigma$200$\omega$180) lead to strong effects of self-interactions, while the smaller cross-sections (SIDM$\sigma$02$\omega\infty$, SIDM$\sigma$20$\omega$180) represent a more conservative scenario that is closer to CDM behaviour (see Fig.~\ref{fig0}). Nonetheless, on large scales ($\gtrsim 10 \,\Mpch$), the effects of self-interactions are negligible, and SIDM behaves similarly to CDM. Using these cross-sections allows us to explore the impact of self-interactions on the topology of the cosmic web and to assess how sensitive our results are to the strength of these interactions.

In this work, we analysed snapshots at $z=2$ and $z=0$. 
The halos were extracted with the \textsc{SUBFIND} algorithm implemented in the code \textsc{OpenGadget3} \citep{springel_2001, dolag_2009}. 
At redshift $z=0$, the virial radius of halos ranges from $0.01\Mpch$ to $1\Mpch$ while the mass ranges from dwarf galaxy masses ($10^{8}~\mathrm{M_\odot}$) to galaxy cluster masses ($10^{14}~\mathrm{M_\odot}$).  At redshift $z=2$, the upper bound of the halo mass range decreases to $10^{13}~\mathrm{M_\odot}$, while the lower bound remains unchanged.
The details of the analysis are described next.

\subsection{Analysis of the cosmic web structure}

We use the genus to compare the topology of the cosmic web in CDM and SIDM cosmologies. The genus quantifies the connectivity of the surface densities at different threshold values. It is one of the Minkowski functionals, and it has been introduced in large-scale structure studies by \cite{gott_quantitative_1987,mecke_robust_1993}. 

\subsubsection{Estimation of the density}

A Delaunay tessellation \citep{delaunay_sur_1934} is first built \citep[with the \textsc{DisPerSE} implementation,][]{sousbie_persistent_2011} on the discrete distribution of DM particles, and the density of each tetrahedron is taken as its inverse volume. The tessellation is then projected onto a regular grid of $1000^3$ pixels of size $0.048\Mpch$. This approaches avoid regions of the box with artificially null density due to the discrete distribution of tracers. 

The density field is then smoothed  with a $3$D Gaussian kernel of width $\RG$. 
We then define the density contrast:
\begin{equation}
    \nu = \frac{\rho - \bar{\rho}}{\sigma}\,,
\end{equation} 
where $\rho$ is the smoothed density, $\bar{\rho}$ the mean density in the simulated box and $\sigma$ the standard deviation of the density distribution. 

Besides estimating the genus on the full box, we also estimate it in boxes centred around individual halos, of size 3 times their virial radii.

\subsubsection{Measurement of the genus}

For a given $\nu$ threshold, an isodensity surface separating the region with higher density from the region with lower density can be defined in the 3D space. The genus is a measurement of the connectivity of the isosurface. In practice, we parametrise the curve with the volume-fraction threshold $\nuf$. For a given value of $\nu$, $\nuf$ is the density threshold for which the isodensity surface encloses the same volume fraction than in a Gaussian field at the same smoothing length, as commonly done in the literature \citep[e.g.][]{Codis2013,appleby_cosmological_2018}. For example, when $\nuf=0$, the genus will be computed for the density contrast which separates in equal volumes the density field above and below. This parametrization allows to eliminate  the non-gaussianity in the 1-point distribution of the field values \citep{gott_quantitative_1987, weinberg87, melott88}. When parametrised in this way, differences in the genus curves in CDM and SIDM will be solely driven by different topologies (connectivity of the isosurfaces). The Figure in Appendix~\ref{sec:filling_factor} illustrates the dependence of the volume-fraction threshold $\nuf(\nu)$ on the density contrast. As one increases the smoothing scale $\RG$, the simulated density contrast gets closer to Gaussian, and therefore $\nuf(\nu)$ gets closer to identity. As the smoothing scale decreases, the density contrast $\nu$ is very skewed towards high density values. We also note that due to the the discretisation of the density field and the limited volume on which the genus is computed, many pixels in voids are assigned to the same low-density value. Therefore one cannot probe small values of $\nuf$, especially when the genus is computed on small boxes. This effect is smeared out when the field is smoothed with larger $\RG$.

\begin{figure}
    \centering
    \includegraphics[width=\linewidth]{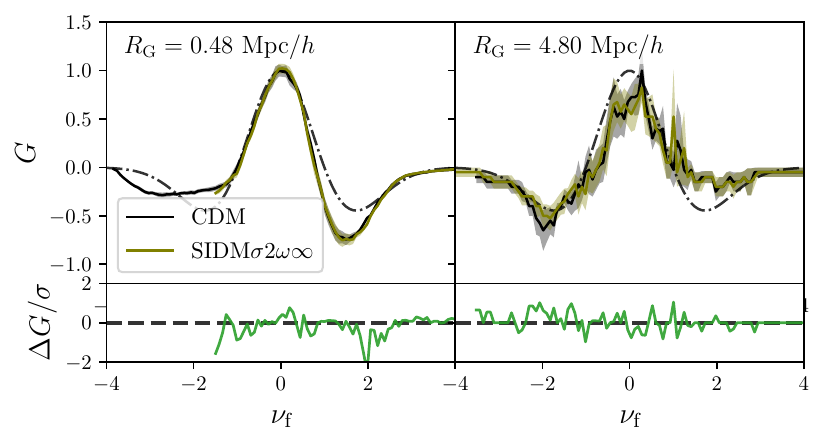}
    \caption{Upper panels: Normalised genus of  the full $48\Mpch$ cosmological box as a function of the volume-fraction threshold, $\nu_f$, for CDM (black) and SIDM (green, $\sigma_0=2~\rm{cm}^2/\rm{g},\,\omega_0=\infty$) at $z=0$, at two smoothing scales, $\RG=0.48$ (left) and $4.8\Mpch$ (right). The shaded area is computed from the standard error on the mean genus computed over 8 independent sub-boxes. The black dash-dotted curve shows the analytical prediction for a Gaussian random field in three dimensions (normalised to the same maximum value than the simulation). 
   Lower panels: residuals between the CDM and SIDM genus curves normalised by the standard mean error of the calculation on sub-boxes. }
    \label{fig:genusFullbox}
\end{figure}
\subsubsection{Genus statistics}

\begin{figure}
    \centering
    \includegraphics[width=0.9\linewidth]{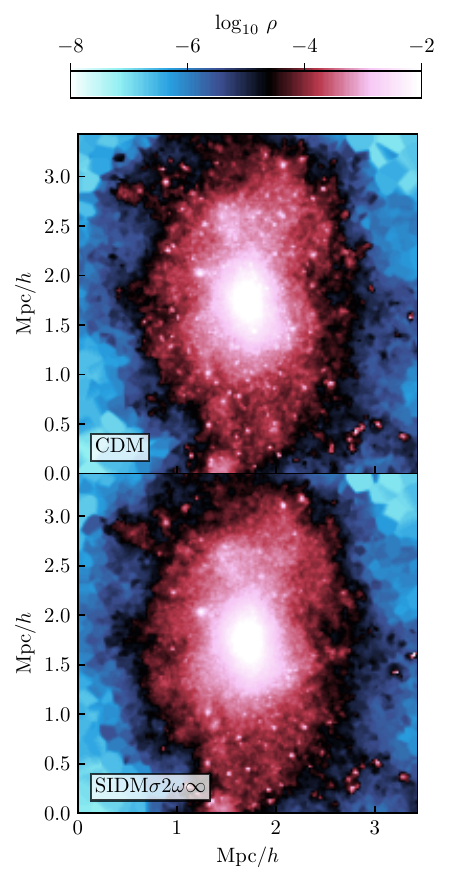}
    \caption{Dark matter density field, $\rho$, interpolated on a Delaunay tessellation in a volume of size $3.4\Mpch$ centred on the most massive halo of the simulation ($M_{\rm vir}=1.3\times10^{14}\Msolh$, $R_{\rm vir}=1.1\Mpch$), for CDM (top) and SIDM ($\sigma_0=2~\rm{cm}^2/\rm{g}$ and $\omega_0=\infty$) (bottom). The density fields are shown in logarithmic scale and averaged over $\sim 100$~kpc.
    }
    \label{fig:zoomHalos}
\end{figure}

The physical properties of the large-scale structure can be studied through the topology of such surface density \citep{gott_quantitative_1987, park_topology_2005, watts_large-scale_2017}. Minkowski functionals are mathematical tools \citep{minkowski_volumen_1903}, based on integral geometry that have been used in cosmology to characterise the large-scale structures through its connectivity, shape and content \citep[see][]{mecke_robust_1993,liu_cosmological_2025,armijo_cosmological_2025}. 

The genus is defined by: 
\begin{equation}
\label{eq:genus}
    G = \text{nb. of holes } - \text{ nb. of isolated regions.} 
\end{equation}

In this definition, the genus is equal to minus one half of the Euler characteristics $\chi$,  $G=-\chi/2$. Alternatively, one can figure the genus as the number of cuts through a surface required to separate it into two pieces. In topology, the meaning of a hole is similar to that of a torus. A high number of holes means that the topology of the surface density is "sponge-like", that is, the connectivity is high, therefore the genus $G$ is positive. The genus becomes negative when the number of closed surfaces is higher than the number of holes, and we call the distribution "meatball like". It is now clear that the genus depends on the density threshold adopted. At low density threshold $\nu$, the surface density draws many closed surfaces, enclosing low density regions: voids. On the other hand, at high density threshold, the surface density encloses dense regions: clumps. In both cases, the corresponding genus is negative. Then, at intermediate density threshold, close to the mean density value, the field is highly connected (sponge-like), consequently, the surface has many holes, leading to a positive genus.

The genus as a function of the density contrast $g(\nu)$ for a Gaussian random field is known \citep{doroshkevich_spatial_1973, adler_geometry_1981, hamilton_topology_1986}: 
\begin{equation}
    g(\nu)\propto (1-\nu^2) \, \e^{-\nu^2/2} \, .
\end{equation}

This function exhibits a symmetry in the density as $g(\nu)=g(-\nu)$, which is expected for a Gaussian field, i.e the level of connectivity should be equal in high density and low density regions. At infinitely high or low density, the genus goes to zero, as there are no isodensity surfaces. When computing the genus on the box, we divide the box in eight sub-boxes and compute the genus individually, then we can use the mean value and the dispersion from it of the eight boxes to interpret the results.

Statistical metrics have been developed from the genus, and we use some of them to better quantify the differences found in the genus of CDM and SIDM. In particular, we use the clump abundance $A_\mathrm{C}$ defined by \cite{park_topology_2001,park_topology_2005,park_effects_2005} as:
\begin{equation}
    \label{clumpAb}
    A_\mathrm{C}=\int G \,\mathrm{d}\nu_{\rm f} \, ,
\end{equation}
which represents the area under the genus curve on a specific integration range. We chose the edges of the integral to be $1.2<\nu_{\rm f}<2.2$, following previous definitions in the literature: the interval is centred near the position of the minima of the Gaussian genus ($\nu_{\rm min}\sim \sqrt{3}$), and is sufficiently high to minimize a bias due to an eventual shift of the genus curve \citep{park_topology_2005}. The genus at density contrast larger than $\nu_{\rm min}$ would essentially correspond to counting clumps, since the field at these densities has a  meatball-like topology. However, our integration range  $1.2<\nu_{\rm f}<2.2$  is designed to characterize the topology at density contrasts where the transition from a percolated network to a clump-dominated topology occurs (\citet{2000PhRvL..85.5515C} found the percolation level to be around $\nu_f=1.5$, i.e close to the minimum of genus). In this sense, our metric quantifies the connectivity of the field and is complementary to more classical statistics like peak count.

To compare the result in both models, we refer to the ratio of clump abundances $A_{C, \rm CDM}/A_{C,\rm SIDM}$. The value of this ratio is interpreted as follows: if $A_{C, \rm CDM}/A_{C,\rm SIDM} > 1$, it means that the genus in CDM is more negative than in SIDM in the overdensity region, meaning that CDM has more isolated high-density regions than SIDM, i.e. CDM is more clumpy than SIDM. In contrast, if $A_{C, \rm CDM}/A_{C,\rm SIDM} < 1$, it means that SIDM is more clumpy than CDM. Therefore, this ratio quantifies how much CDM and SIDM differ in terms of clumpiness. The higher the ratio, the more discrepancies there are between the two models. If the ratio is equal to one, it means that both models have the same clump abundance, i.e. neither CDM nor SIDM have a clumpier distribution than the other. We expect $A_{C, \rm CDM}/A_{C,\rm SIDM} \geq 1$, indicating that CDM is able to form more clumps than SIDM due to the absence of self-interactions. The self-interactions in SIDM redistribute energy among the particles. Therefore, it is more difficult for SIDM to form dense clumps, leading to a smoother distribution compared to CDM.

\subsection{Getting closer to the observations}
Galaxy surveys such as those conducted by Euclid \citep{collaboration_euclid_2025} measure the density field of galaxies in a two-dimensional space in a thin redshift bin. Moreover, the DM density field is not directly observable but is reconstructed from the measured galaxy density field. Genus measurements are applied directly on the distribution of galaxies that are hosted in DM halos. Therefore, in order to produce results comparable to such observations, we applied the genus analysis directly on the halo density field. To do so, we estimate the density field from the two-dimensional projected positions of the halos. Each halo is therefore treated as a point-like object, weighted by its virial mass. In two dimensions, the interpretation of the genus differs from the three-dimensional case \citep{melott88}. The two-dimensional genus is defined as
\begin{equation}
\label{eq:genus2D}
\begin{split}
    G_{\rm 2D} =\ & \text{nb. of isolated high-density regions} \\
    &- \text{nb. of isolated low-density regions.}
\end{split}
\end{equation}
For a two-dimensional Gaussian field, we expect the genus to be negative for negative density contrast values and positive for positive density contrast values. In this case, the genus is given by:
\begin{equation}
    G_{\rm 2D}\propto \nu \, \e^{-\nu^2/2}\, .
\end{equation}
The genus analysis is carried out in the same way as described for the three-dimensional case. We use the clump abundance $A_{\rm C}$ which has the same definition but with an interval shifted to $0.5<\nuf<1.5$, in order to be centred around the extremum of the two-dimensional genus of a Gaussian field ($\nu_{\rm min}=1$).

\section{Results}
\label{Results}

This section presents the results of our comparison between the two DM models. We first apply the genus statistic to the full simulation box. We then extend the analysis to subregions centred on DM halos, where we use the clump abundance ratio to quantify the differences between the models. Finally, we investigate the dependence of the signal on the smoothing scale and redshift before considering the two-dimensional configuration that better reflects observational data.

\begin{figure}
    \centering
    \includegraphics[width=1\linewidth]{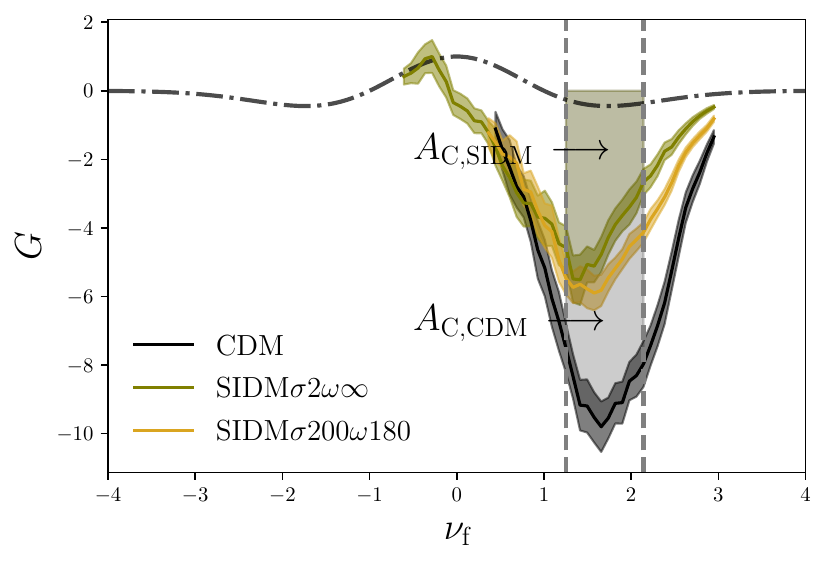}
    \caption{Genus curves of the density field smoothed at $\RG=0.01\Mpch$ within the sub-box centred on the most massive halo (whose density is shown in Figure~\ref{fig:zoomHalos}). The genus is displayed for CDM (black) and two SIDM models (yellow and green). 
    For reference, the black dash-dotted curve shows the analytical genus expected for a three-dimensional Gaussian random field with ad-hoc normalisation (Eq.~\eqref{eq:genus}). The vertical shaded regions highlight the intervals used to compute the clump abundance parameter, $A_{\rm C}$, which is centred \textbf{near} the minimum of the genus curve on the gaussian random field. }
    \label{fig:genusHaloZoom}
\end{figure}

\subsection{CDM and SIDM genus comparison on large scales}

Figure~\ref{fig:genusFullbox} presents the genus computed on the entire $48\Mpch$ box in the CDM (black) and SIDM (green) models at two different smoothing scales, $\RG=0.48$ and $4.8\Mpch$ (left and right panels, respectively). The genus curves of both models are normalised here by their maximum amplitude. Uncertainties are derived from the standard error on the mean genus computed over 8 independent sub-boxes.
For comparison, we also display the genus curve of a gaussian random field smoothed at the same scale and normalised so that its maximum matches the one on the simulated data. 
We note that both CDM and SIDM models show a clear departure from gaussianity, particularly at a small smoothing scale, where the genus curves become increasingly asymmetric.
At $\RG=0.48\Mpch$ and at high $\nu_{\rm f}$ thresholds, the density field therefore exhibits a characteristic ``meatball-like'' topology, where the matter distribution is dominated by numerous disconnected clumps corresponding to DM halos hosting galaxies. In contrast, increasing the smoothing scale reduces the number of isolated structures and drives the topology towards a more Gaussian-like regime. 

Despite these non-Gaussian features, the genus curves of CDM and SIDM remain remarkably similar for both smoothing scales. 
To quantify the differences, we compute the residuals between the CDM and SIDM genus curves normalised by the standard mean error estimated from the sub-boxes ($\sigma=\sqrt{\sigma_{\rm G_{SIDM}}^2+\sigma_{\rm G_{CDM}}^2}$). In all cases, the residuals remain within the $3\sigma$ level, indicating that no statistically significant difference is detected between the large-scale topologies of CDM and SIDM.  
This was expected from \cite{2020MNRAS.497.3809S, fischer_cosmological_2022}, where the CDM and SIDM density fields where found very similar at large scale, consistently with the fact that SIDM is designed to preserve the successful large-scale predictions of CDM while primarily modifying the inner structure of halos. 
This naturally motivates a shift towards smaller scales, where SIDM effects are expected to be stronger.

\subsection{Genus around the most massive halo at small scale}
\label{aroundHalos}

We now move to characterizing the small-scale topology of the density field with higher sensitivity, by focusing on the density field around individual massive DM halos. 

To motivate this approach, Fig.~\ref{fig:zoomHalos} shows the density field around the most massive halo of the simulation for both CDM and SIDM. Although the overall large-scale morphology remains similar, the density field is visually clumpier in CDM than in SIDM. Our objective is to explore whether the genus is relevant to statistically capture this difference. We focus first on the region around the most massive halo, where we expect to detect the largest discrepancy between CDM and SIDM models. For this halo, we then measure the 3D genus curves in a box of size $L_{\rm box}=3R_{\rm vir}$, where $R_{\rm vir}$ is the virial radius of the halo. This choice ensures that the analysis captures the halo and its surrounding structures, where SIDM self-interactions are expected to produce the strongest modifications of the density distribution and topology. We also repeated the analysis on larger regions around the halos and found that for regions larger than $L_{\rm box}\gtrsim10 R_{\rm vir}$, the genus no longer significantly differentiates the topology of the two DM models, because the signal present in the immediate vicinity of the halo becomes diluted with the rest of the box. However, lowering the box size diminishes the signal-to-noise ratio since the genus is measured over a smaller number of volume elements. We found that a box size of $3R_{\rm vir}$ offers the best compromise in terms of signal and signal-to-noise ratio.

Figure~\ref{fig:genusHaloZoom} shows the  genus curve at the smallest smoothing scale ($\RG=0.01\Mpch$). We note significant differences between CDM and SIDM at high density thresholds ($1 \lesssim \nuf \lesssim 3$). In this regime, the CDM genus reaches substantially more negative values than with the SIDM models, with discrepancies up to $\sim8\sigma$. Since negative genus values at high $\nuf$ trace isolated overdense structures, this result indicates that the CDM density field contains a significantly larger population of compact high-density clumps, consistent with the visually clumpier substructure distribution seen in Fig.~\ref{fig:zoomHalos}.
The velocity-dependent SIDM model (yellow curve) exhibits slightly more negative genus values than the constant cross-section SIDM model (green curve), corresponding to a somewhat clumpier topology. This behaviour is expected in massive halos, where the large velocity dispersion suppresses the effective self-interaction cross-section in velocity-dependent SIDM models (see Fig.~\ref{fig0}). As a consequence, self-interactions become less efficient at erasing substructures, allowing a larger number of dense clumps to survive compared to the constant cross-section case.

\begin{figure}
    \centering
    \includegraphics[width=1\linewidth]{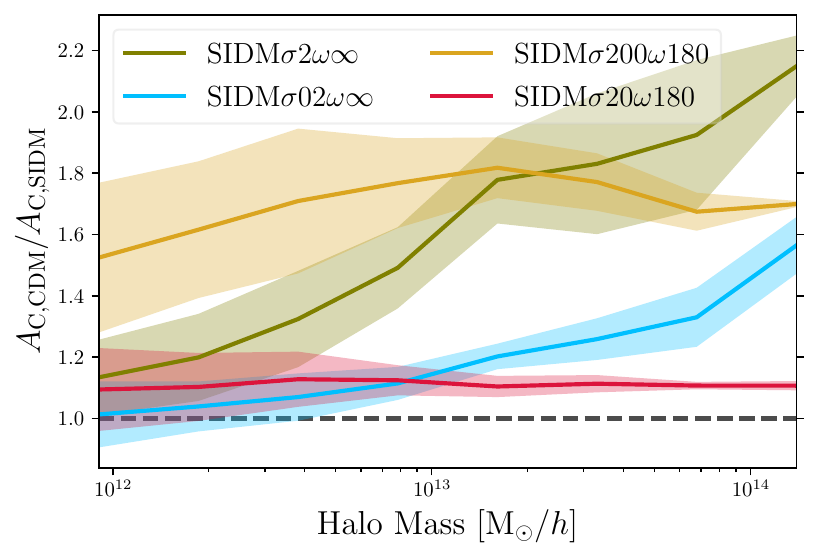}
    \caption{Ratio of the clump abundance parameter (excess of isolated high-density structures, defined in Eq.~\ref{clumpAb}) at $\RG=0.01\Mpch$ between CDM and the four SIDM models, $A_{\rm C,CDM}/A_{\rm C,SIDM}$, measured at $z=0$ as a function of halo mass. Uncertainties are derived from bootstraping. The analysis is performed on a sample of 338 halos cross-matched across the five DM simulations using their masses and spatial positions. Ratios larger than unity indicate that CDM halos exhibit a clumpier topology than their SIDM counterparts.}
    \label{fig:AC-allSIDM}
\end{figure}

\subsection{Generalization at all massive halos: clump abundance ratio from the genus curves}

To systematically quantify the differences between the genus highlighted in Fig.~\ref{fig:genusHaloZoom} for the most massive halo, we use the clump abundance parameter, $A_{\rm C}$, which quantifies the area enclosed below the genus curve in the high-density regime (Eq.~\eqref{clumpAb}).
The shaded regions in Fig.~\ref{fig:genusHaloZoom} illustrate the intervals used to compute $A_{\rm C}$. For the most massive halo shown in Fig.~\ref{fig:zoomHalos}, and using a smoothing scale of $\RG=0.01\Mpch$, we measure a clump abundance ratio of $A_{\rm C} \simeq 2$, indicating that the CDM density field is approximately twice as clumpy as its SIDM counterpart in the vicinity of this halo. 

We generalised this analysis to the whole halo sample by considering the 338 halos with masses between $10^{12}$ and $10^{14} \Msol$ cross-matched across the five DM simulations using their masses and spatial positions. For each halo, we compute the genus within a sub-box of size $L_{\rm box}=3R_{\rm vir}$ centred on the halo and evaluate the clump abundance on density fields smoothed at $\RG=0.01\Mpch$. Figure~\ref{fig:AC-allSIDM} shows the resulting clump abundance ratios as a function of the halo mass for the different SIDM models. The solid curves represent the mean ratios measured in mass bins, while the shaded regions correspond to bootstrap uncertainties obtained from 1000 resamplings of the halo population, with the uncertainties on the genus curve around each halo (measured over the 8 sub-boxes) added in quadrature. For completeness, we provide in Appendix~\ref{sec:clump_abundance} the individual halo measurements used to derive the average curve shown in Fig.~\ref{fig:AC-allSIDM}, together with their associated uncertainties.

A clear trend emerges with increasing self-interaction cross-section. For constant cross-section SIDM models, the clump abundance ratio systematically increases with halo mass, reaching values above $A_{\rm C} \sim 2$ for the most massive halos in the $\sigma_0=2\;\rm{cm}^2\,\rm{g}^{-1}$ model. This behaviour indicates that self-interactions efficiently suppress small-scale substructures and smooth the inner halo density field, leading to significantly less clumpy topologies than in CDM. In comparison, the SIDM model with $\sigma_0=0.2\;\rm{cm}^2\,\rm{g}^{-1}$ remains closer to the CDM prediction, even though it is significantly higher than unity for the highest halo masses. Introducing a velocity-dependent cross-section significantly modifies this trend. In massive halos, where the velocity dispersion is high, the effective SIDM cross-section is strongly reduced (see Fig.~\ref{fig0}). As a consequence, self-interactions become less efficient at dynamically heating and erasing substructures, allowing dense clumps to survive more easily. This effect is illustrated by the velocity-dependent SIDM model with $\sigma_0=200\,\rm{cm}^2 \, \rm{g}^{-1}$ and $\omega=180\,\rm{km} \,\rm{s}^{-1}$ (yellow curve in Fig.~\ref{fig:AC-allSIDM}), for which the clump abundance ratio flattens and even decreases towards high halo masses. The differences from the CDM case detected in the topology of these massive halos therefore cease to increase, despite the large self-interaction cross-section in the low-velocity regime.

\begin{figure}
    \centering
    \includegraphics[width=1\linewidth]{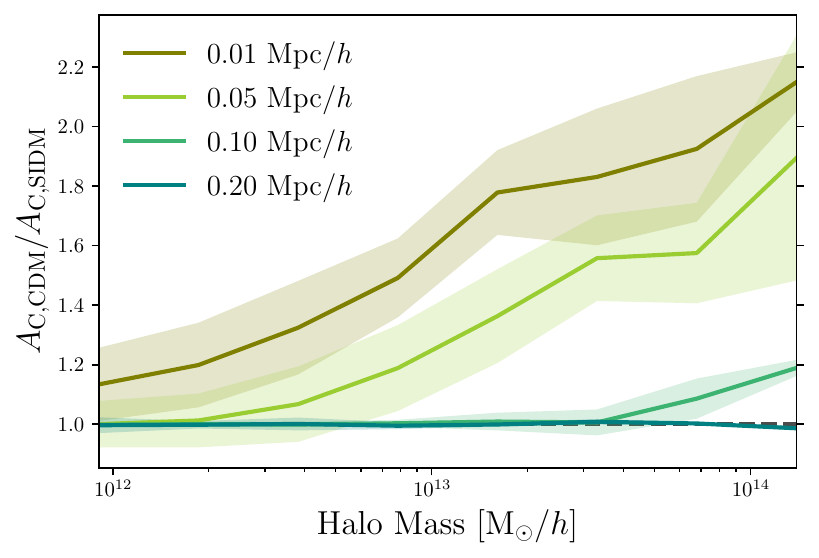}
    \caption{Ratio of the clump abundance parameter, $A_{\rm C}$, between CDM and SIDM ($\sigma_0=2\,\rm{cm}^2/\rm{g}$ and $\omega_0=\infty$) as a function of halo mass for four different smoothing scales of the density field around these halos, $\RG$, measured at $z=0$ for the same sample of halos as Fig.~\ref{fig:AC-allSIDM}.
    }
    \label{fig:clumpAbundance}
\end{figure}

\subsection{Scale-dependency of the signal}

In Figure~\ref{fig:clumpAbundance}, we investigate how the clump abundance ratio varies with the smoothing scale of the density field, comparing the CDM model with the SIDM model with constant cross-section $\sigma_0 = 2\, \rm{cm}^2 \, \rm{g}^{-1}$. Unsurprisingly, the strongest deviations from CDM are found at the smallest smoothing scales, $\RG=0.01$ and $0.05\Mpch$, where the topology is most sensitive to halo substructures. At $\RG=0.01\Mpch$, the clump abundance ratio reaches values as high as $A_{\rm C} \simeq 2.1$ for the most massive halos, indicating that CDM halos are more than twice as clumpy as their SIDM counterparts. This behaviour is fully consistent with the expectation that DM self-interactions suppress small-scale substructures and smooth the inner density field in dense environments. Massive halos appear to be the most discriminating systems for velocity-independent cross-sections. The deviations from CDM are reduced for the most massive systems when considering velocity-dependent cross-sections, as illustrated in Fig.~\ref{fig:clumpAbundance}. At $\RG=0.01\Mpch$, CDM remains significantly clumpier than SIDM down to halo masses of $M \sim 10^{12}\Msol$, while the discrepancy exceeds a factor of two for regions around halos with $M \gtrsim 10^{13}\Msol$. As the smoothing scale increases, the clump abundance ratio progressively approaches unity, showing that the topological differences between CDM and SIDM are primarily driven by small-scale structures. Correspondingly, the minimum halo mass for which a significant difference is detected also increases with $\RG$. For smoothing scales larger than $\RG \gtrsim 0.1\Mpch$, the genus curves become nearly indistinguishable and no significant difference between CDM and SIDM is detected.

\begin{figure}
    \centering
    \includegraphics[width=1\linewidth]{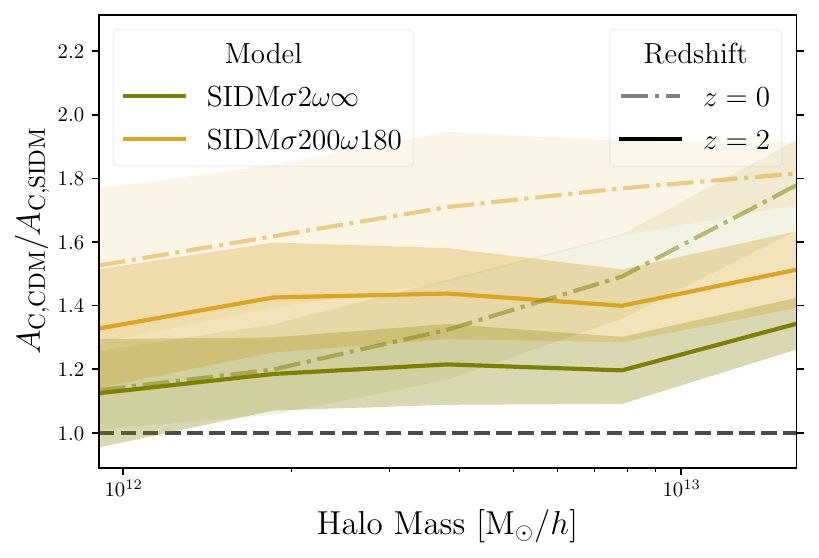}
    \caption{Clump abundance ratio  measured at $z=2$ (solid lines) and $z=0$ (dash-dotted lines) as a function of halo mass, for a density field around these halos smoothed over $\RG=0.01\Mpch$ between CDM and two SIDM models. The analysis is based on a sample of 220 halos cross-matched across the three DM simulations using their masses and spatial positions.}
    \label{fig:clumpAbundancez2vs0}
\end{figure}

\subsection{Redshift evolution of the signal}

The topology of the large-scale structure evolves over cosmic time as gravitational collapse progressively amplifies density fluctuations and builds increasingly complex nonlinear structures. To investigate how the topological differences between CDM and SIDM evolve with redshift, we repeat the previous analysis at $z=2$ and compare the resulting clump abundance ratios, $A_{\rm C}$, between the different DM models. Figure~\ref{fig:clumpAbundancez2vs0} presents the clump abundance ratio at $z=2$ for the two SIDM models exhibiting the strongest deviations from CDM. In both cases, even though the ratios are still significantly higher than unity, they are systematically smaller than those measured at $z=0$, indicating that the topological differences between CDM and SIDM are weaker at earlier cosmic times. This behaviour is expected because self-interactions have less time to dynamically heat halos, redistribute matter, and suppress small-scale substructures at high redshift. Moreover, the formation of structures is in a less advanced stage, with fewer massive structures formed, even in the CDM scenario. Nevertheless, the dependence of the clump abundance ratio on halo mass remains qualitatively similar to the trend observed at $z=0$, with more massive halos exhibiting stronger deviations from the CDM prediction. Finally, our results suggest that low-redshift massive structures provide the most favorable environments to maximise the detectability of SIDM topological signatures.

\begin{figure}
    \centering
    \includegraphics[width=0.9\linewidth]{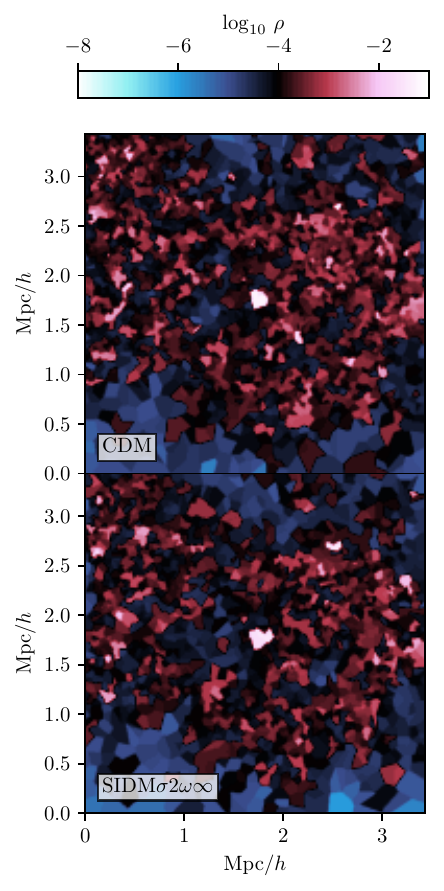}
    \caption{Two-dimensional projected density field of DM halos centred on the most massive halo of the simulation, corresponding to the same region shown in Fig.~\ref{fig:zoomHalos}, but with the density directly estimated in 2D in a thick tomographic slice ($48$~Mpc$/h$). The top and bottom panels show the projected halo density field for CDM and the SIDM model with $\sigma_0=2\,\rm{cm}^2 \, \rm{g}^{-1}$ and $\omega_0=\infty$ respectively.}
    \label{fig:density2D}
\end{figure}

\begin{figure}
    \centering
    \includegraphics[width=\linewidth]{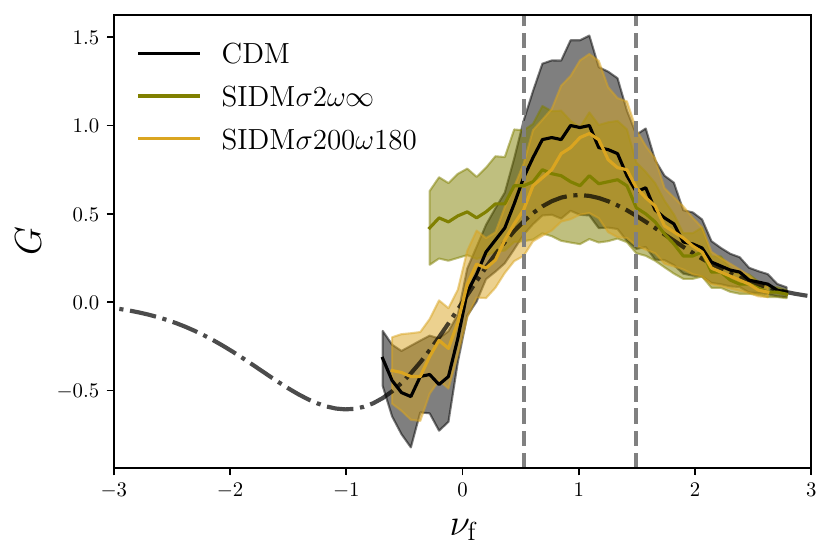}
    \caption{Two-dimensional genus computed from the projected halo density field shown in Fig.~\ref{fig:density2D}, smoothed at $\RG=0.01\Mpch$. The genus curves are displayed for CDM (black) and two SIDM models with different self-interaction cross-sections. The black dash-dotted curve represents the analytical genus expected for a two-dimensional Gaussian random field. Shaded regions indicate the scatter estimated from the sub-box analysis, while the vertical grey dashed lines delimit the integration interval used to compute the clump abundance parameter.}
    \label{fig:genusHaloDensityField}
\end{figure}

\subsection{Forecast for observations: 2D genus from the halo density field}

In modern observational surveys, the density field is often estimated directly from the galaxy distribution. Our results suggest that the strongest discrepancy between CDM models occurs at a small smoothing scale of the density field ($\RG < 0.1\Mpch$). To reach such a small scale, one needs to estimate the density field with a large enough number of tracers, which is not possible with spectroscopic sample \citep[where the mean galaxy separation is usually larger than $\sim$1~Mpc, see e.g. figure~1 of][for the \textit{Euclid} spectroscopic survey]{kraljic2026}. Deep photometric sample offers a promising alternative, at the price of a low precision of galaxy position along the line-of-sight (with a photometric redshift uncertainty leading usually to a confusion $>50$~Mpc$/h$ along the line-of-sight). To get a configuration as close as possible from the observations, we investigate  whether the genus remains sensitive to the differences between CDM and SIDM when computed in 2D on the halo distribution rather than on the full DM particle density field.

The resulting genus curves computed from the projected halo density field are shown in Fig.~\ref{fig:genusHaloDensityField}. Overall, the genus measured from the halo distribution exhibits qualitative trends similar to those obtained from the full DM density field, although the signal is significantly noisier due to the much smaller number of tracers. At high density thresholds ($1 \lesssim \nuf \lesssim 3$), the CDM genus reaches systematically larger positive values than the SIDM models, indicating a larger abundance of isolated high-density structures in the projected halo distribution \citep{1989ApJ...345..618M}. This result is fully consistent with the visual impression from Fig.~\ref{fig:density2D}, where the CDM halo field appears substantially clumpier than its SIDM counterpart. The difference is particularly pronounced for the SIDM model with constant cross-section, whose smoother topology reflects the suppression of small-scale halo substructures induced by DM self-interactions (see Fig.~\ref{fig:genusHaloDensityField}). In contrast, the velocity-dependent SIDM model remains closer to the CDM prediction, consistent with the reduction of the effective cross-section in high-velocity environments. 

Interestingly, the genus curves also show that SIDM with constant cross-section exhibits a systematically higher genus than CDM for $\nuf < 0.5$ (see Fig.~\ref{fig:genusHaloDensityField}). In the context of the two-dimensional genus definition, this behaviour suggests that the SIDM halo distribution contains fewer isolated clumps and forms a more connected network of isodensity contours than CDM. Such a trend is qualitatively consistent with the expected small-scale substructure suppression effect induced by self-interactions. However, this feature was not clearly observed in the genus analysis performed on the DM particle density field of the entire 48 $\Mpch$ size box.

\begin{figure}
    \centering
    \includegraphics[width=\linewidth]{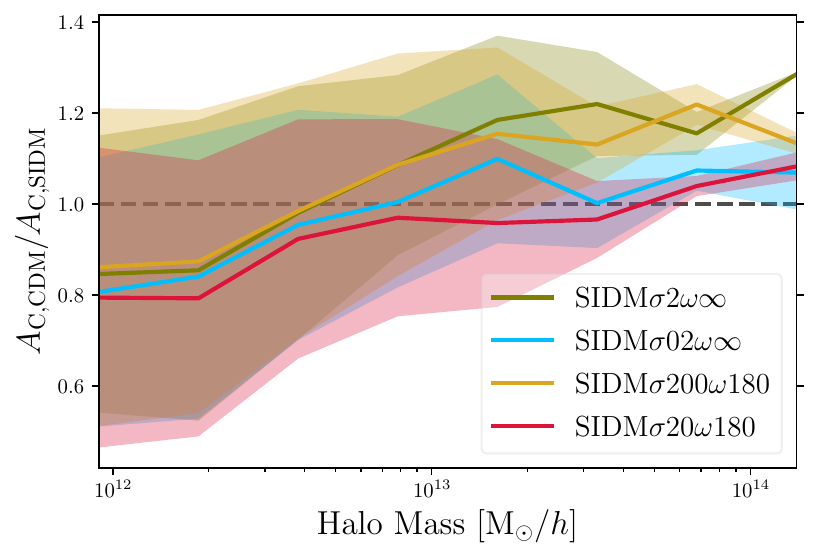}
    \caption{Same as in Fig.~\ref{fig:AC-allSIDM} but for the genus computed from the projected halo density field at $\RG=0.01\Mpch$. The analysis is performed on regions centred around 325 halos for the SIDM ($\sigma_0=2\,\rm{cm}^2 \, \rm{g}^{-1}$ and $\omega_0=\infty$) model and 327 halos for the SIDM ($\sigma_0=2\,\rm{cm}^2 \, \rm{g}^{-1}$ and $\omega_0=180\,\rm{km} \, \rm{s}^{-1}$) model.}
    \label{fig:placeholder}
\end{figure}

Figure~\ref{fig:placeholder} shows that the two-dimensional genus analysis of the projected halo density field are broadly consistent with those previously derived from the three-dimensional DM density field. The constant cross-section SIDM model with $\sigma_0=2\,\rm{cm}^2\,\rm{g}^{-1}$ again shows the strongest suppression of small-scale structures, while models with weaker or velocity-dependent cross-sections remain closer to the CDM prediction. However, the differences measured from the projected halo distribution are generally weaker than those obtained from the full three-dimensional DM density field. In the 3D analysis, the clump abundance ratio reaches values as high as $A_{\rm C} \sim 2.2$ for the most massive halos, whereas the projected halo analysis typically yields smaller deviations and larger uncertainties (see Figs.~\ref{fig:clumpAbundancef2vsf02} and ~\ref{fig:placeholder}). This reduction in sensitivity is expected for several reasons. First, the halo distribution provides a much sparser sampling of the underlying matter field than the full DM particle distribution, leading to noisier genus measurements. Second, the projection from three dimensions to two dimensions partially washes out the small-scale topological features associated with compact substructures.

Despite these limitations, the projected halo genus preserves the main SIDM signatures identified in the full three-dimensional DM density field, with the clump abundance ratio remaining systematically above unity for the strongest SIDM models, particularly around massive halos (see Fig.~\ref{fig:placeholder}). This demonstrates that genus measurements performed on halo or galaxy distributions remain sensitive to DM self-interactions even when using observable-like tracers instead of the full DM particle field. These results are therefore encouraging for future applications to galaxy surveys such as \textit{Euclid}, although several observational effects and systematic uncertainties will need to be carefully investigated, including projection effects, sparse sampling, and the differences between two-dimensional observational reconstructions on baryonic matter (galaxies) and the intrinsic three-dimensional dark matter distribution.

\section*{Conclusion and discussion}
\addcontentsline{toc}{section}{Conclusion and discussion}
\label{Conclusion}

In this work, we distinguish for the first time SIDM from CDM through a topological analysis of the density field. We used the genus statistic to evaluate the topology of isodensity surfaces in the different DM models. Compared to other measures, the genus has the advantage to be non-parametric, as it is computed directly from the density field and does not require to run a halo finder, which is inevitably dependent on prior on the definition of a halo. We quantified the connectivity of the density field at high density contrasts through the integrated genus at high density threshold, referred as the clump abundance, $A_{\rm{C}}$. These compared analysis were performed on five Darkium DM-only cosmological simulations: one CDM, two SIDM models with velocity-dependent cross-sections, and two SIDM with velocity-independent cross-sections. Although the genus computed on the entire 48 Mpc/$h$ simulation box does not allow to detect differences between SIDM and CDM, we applied the genus on selected regions centred around halos where self-interactions are expected to have the most significant impact on matter distribution. The region sizes scale 3 virial radii, a scale at which differences can still be seen between the DM theories. We show that CDM is significantly clumpier than SIDM on scales lower than $0.1~\mathrm{Mpc}/h$, as it forms a systematically higher number of isolated high-density regions than found for SIDM. These scales are larger than that of the galactic scale; therefore, the effect of baryons on the topology of the density field is expected to be reduced. This allows us to alleviate the degeneracy with baryonic physics, focusing only on DM. We refer to the clump abundance ratio between CDM and SIDM to quantify the differences between the DM theories. As expected, this ratio is very sensitive to DM self-interactions, characterised by the cross-section. Higher cross-sections allow for more self-collisions and redistribute more energy into DM particles, which prevent the collapse of DM into isolated clumps. This results in a larger genus at high density threshold and lower clump abundance while weaker cross-sections lead to genus curves more similar to that of CDM. These topological differences are more pronounced around halos with mass $>10^{12}~\rm{M}_\odot/h$ at $z=0$. The differences are still visible at higher redshift, $z=2$, although reduced because the collapsed DM structures are in a less advanced stage at this redshift. In order to explore a more observation-like configuration and to assess the detectability of such signal in observations, we computed the genus of the density field interpolated directly from the halo distribution. Since galaxies are expected to form in the centre of DM halos, the distribution of halos traces the underlying DM distribution. In this configuration, the genus still allows one to detect differences between CDM and SIDM, suggesting that the measurements of the clump abundance in observed structures could therefore provide a new statistical probe to constrain SIDM cross-sections and discriminate between different DM models. In practice, in future observations one should target massive halos to disentangle the different SIDM cross-sections and CDM. Moreover, the use of deep surveys is recommended in order to maximise the number of density tracers and to reduce the uncertainty due to small statistics. Moreover, more work is needed to investigate the effect of observational systematics to assess the detectability of such signal on observed structures. Alternatively, the present topological analysis can be applied to compare other DM theories, \citep[e.g.\ warm or fuzzy DM][]{1980PhRvL..45.1980B,GOODMAN2000103} with CDM. One would expect from the gravitational heating induced by such alternative theories to produce similar effects on the genus.

\section{Data Availability}

The data underlying this article is available through reasonable request to the authors. 

\begin{acknowledgements} 

PB acknowledges funding from the CNES post-doctoral fellowship program. This
work was also supported by CNES, focused on the Euclid mission.
MSF gratefully acknowledges the support of the Alexander von Humboldt Foundation through a Feodor Lynen Research Fellowship.

\end{acknowledgements}

\bibliography{src}

\appendix
\section{Filling factor} \label{sec:filling_factor}

\begin{figure}
    \centering
    \includegraphics[width=1\linewidth]{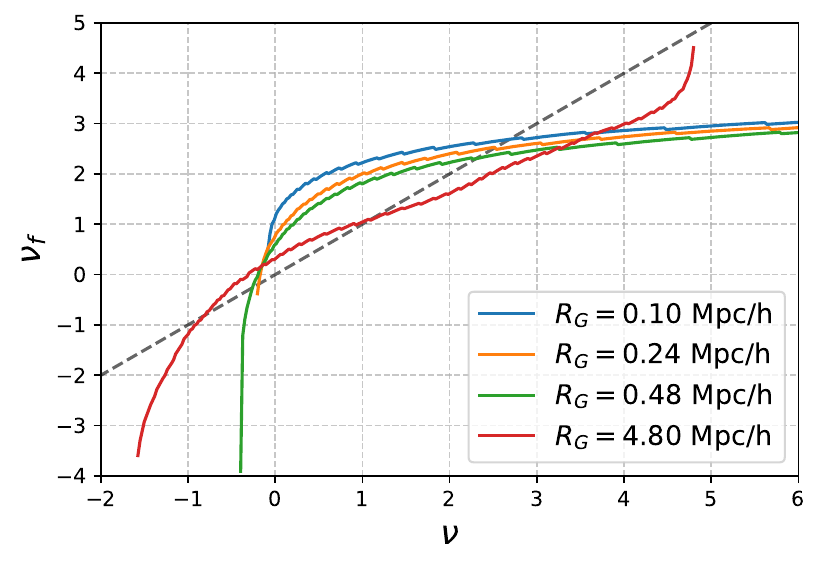}
    \caption{Volume-fraction threshold $\nuf$ as a function of the density contrast $\nu$. The dashed line represents the identity relation $\nuf=\nu$. The volume-fraction threshold is calculated on the full $48\Mpch$ CDM box at redshift $z=0$, smoothed with 4 different Gaussian kernels.}
    \label{fig:nu_f_vs_nu}
\end{figure}

The filling factor is one of Minkowski's functionals, and is defined, for a three-dimensional scalar field, as the fraction of the total volume (or surface for a two-dimensional scalar field) enclosed in surfaces (or contours in two dimensions) as a function of the density contrast. The volume-fraction threshold $\nuf$ is the density contrast corresponding to the same filling factor in a Gaussian field. In Fig.~\ref{fig:nu_f_vs_nu}, we represent the volume-fraction threshold as a function of the density contrast for four smoothing scales (coloured curves) and for a Gaussian field (dashed curve). The interpretation of this figure is included in Sect.~\ref{Method}.

\section{Clump abundance} \label{sec:clump_abundance}

\begin{figure}
    \centering
    \includegraphics[width=1\linewidth]{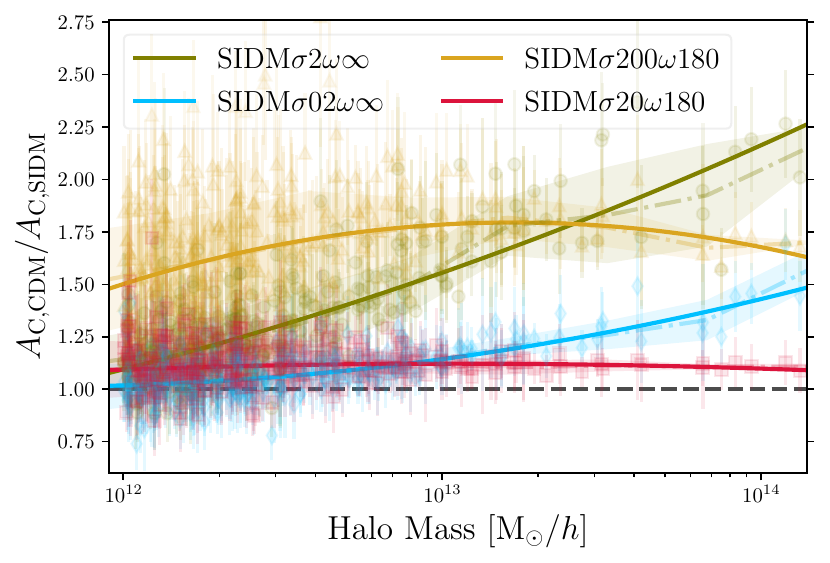}
    \caption{Same as in Fig.~\ref{fig:AC-allSIDM} but additionally showing the individual halo measurements of the clump abundance ratio, $A_{\rm C}$, derived from the genus curves. The error bars represent the propagated uncertainties estimated from the scatter of the genus measured across the 8 sub-boxes. Solid curves show second-degree polynomial fits to the full halo distributions. The dashed curves and shaded regions correspond to the mean relations and bootstrap dispersions in halo mass bins presented in Fig.~\ref{fig:AC-allSIDM}.}
    \label{fig:clumpAbundancef2vsf02}
\end{figure}

The clump abundance ratios presented in Fig.~\ref{fig:AC-allSIDM} are calculated from a discrete sample of halos and evaluated in mass bins equally spaced on a logarithmic scale. This provides clear and readable curves, although information about the scatter of the measured clump abundance is lost. For the sake of completeness, we include in Fig.~\ref{fig:clumpAbundancef2vsf02} all the points, representing individual regions around halos from which the clump abundance is computed. 

\end{document}